\documentclass[12pt]{iopart}
\expandafter\let\csname equation*\endcsname\relax
\expandafter\let\csname endequation*\endcsname\relax

\usepackage{mathtools}
\usepackage{graphics}
\usepackage{graphicx}
\usepackage{dcolumn}
\usepackage{bm}
\usepackage{dsfont} 
\usepackage{amsmath,amssymb}
\usepackage{hyperref}
\usepackage{tabularx}
\usepackage{epsf,epsfig}
\usepackage[normalem]{ulem}
\usepackage[usenames]{color}
\usepackage{multirow}
\usepackage{makecell}
\usepackage{diagbox}
\usepackage[abs]{overpic}
\allowdisplaybreaks
\hypersetup{
    colorlinks=true,
    linkcolor=blue,
    filecolor=magenta,      
    urlcolor=blue,
    citecolor=blue
}
\definecolor{red(ncs)}{rgb}{0.77, 0.01, 0.2}

\usepackage{array}
\newcolumntype{C}[1]{>{\centering\let\newline\\\arraybackslash\hspace{0pt}}m{#1}}

\newcolumntype{C}[1]{>{\centering\arraybackslash}m{#1}}

\begin{document}

\title{Multi-band gravitational wave tests of general relativity}

\author{Zack Carson}
\address{Department of Physics, University of Virginia, Charlottesville, Virginia 22904, USA}
\author{Kent Yagi}
\address{Department of Physics, University of Virginia, Charlottesville, Virginia 22904, USA}

\date{\today}


\begin{abstract}
The violent collisions of black holes provide for excellent test-beds of Einstein's general relativity in the strong/dynamical gravity regime. We  here demonstrate the resolving power one can gain upon the use of multi-band observations of gravitational waves from both ground- and space-based detectors. We find significant improvement in both generic parameterized tests of general relativity and consistency tests of inspiral-merger-ringdown parts of the waveform over single-band detections. Such multi-band observations are crucial for unprecedented probes of e.g. parity-violation in gravity.
\end{abstract}

\maketitle


\label{sec:intro}
\emph{Introduction.}---
Einstein's theory of general relativity (GR) eloquently describes the relationship between the geometries of spacetime and the manifestation of gravity.
After countless observations have held up to the rigors of GR without any sign of deviation, why should we continue to test such a solid theory?
One might argue that while it is impossible to prove a theory is \emph{true}, we can establish constraints on modified theories which may disprove or expand upon our knowledge of gravity.
For example, a more complex theory of gravity could exist in the \emph{extreme gravity} sector where the fields are strong, non-linear, and highly dynamical.
While reducing to the GR we know in the weak gravity limit, such a theory could solidify our understanding of some of the biggest open questions we have: dark energy and the expansion of the universe, dark matter and the galactic rotation curves, inflation in the early universe, or the unification of quantum mechanics and GR.

For over 100 years, GR has been subject to a battery of tests, in search of minute deviations which may indicate alternative theories of gravity.
Countless solar system~\cite{Will_SolarSystemTest}, binary pulsar~\cite{Stairs_BinaryPulsarTest,Wex_BinaryPulsarTest} and cosmological~\cite{Ferreira_CosmologyTest,Clifton_CosmologyTest,Joyce_CosmologyTest,Koyama_CosmologyTest,Salvatelli_CosmologyTest} observations have placed constraints on various modified theories of gravity, all remaining consistent to GR within the noise.
More recently, the observation of gravitational waves (GWs) from the coalescing black holes (BHs) of GW150914~\cite{GW150914} has opened a unique window into gravity, allowing us to probe the extreme gravity sector for the first time~\cite{Abbott_IMRcon2,Yunes_ModifiedPhysics}.
The following 10 binary BH merger events~\cite{GW_Catalogue} and a binary neutron star merger event~\cite{TheLIGOScientific:2017qsa} have similarly identified no significant deviations from Einstein's theory~\cite{Abbott_IMRcon,Monitor:2017mdv,Abbott:2018lct}.

\begin{figure}[htb]
\begin{center}
\includegraphics[width=.65\linewidth]{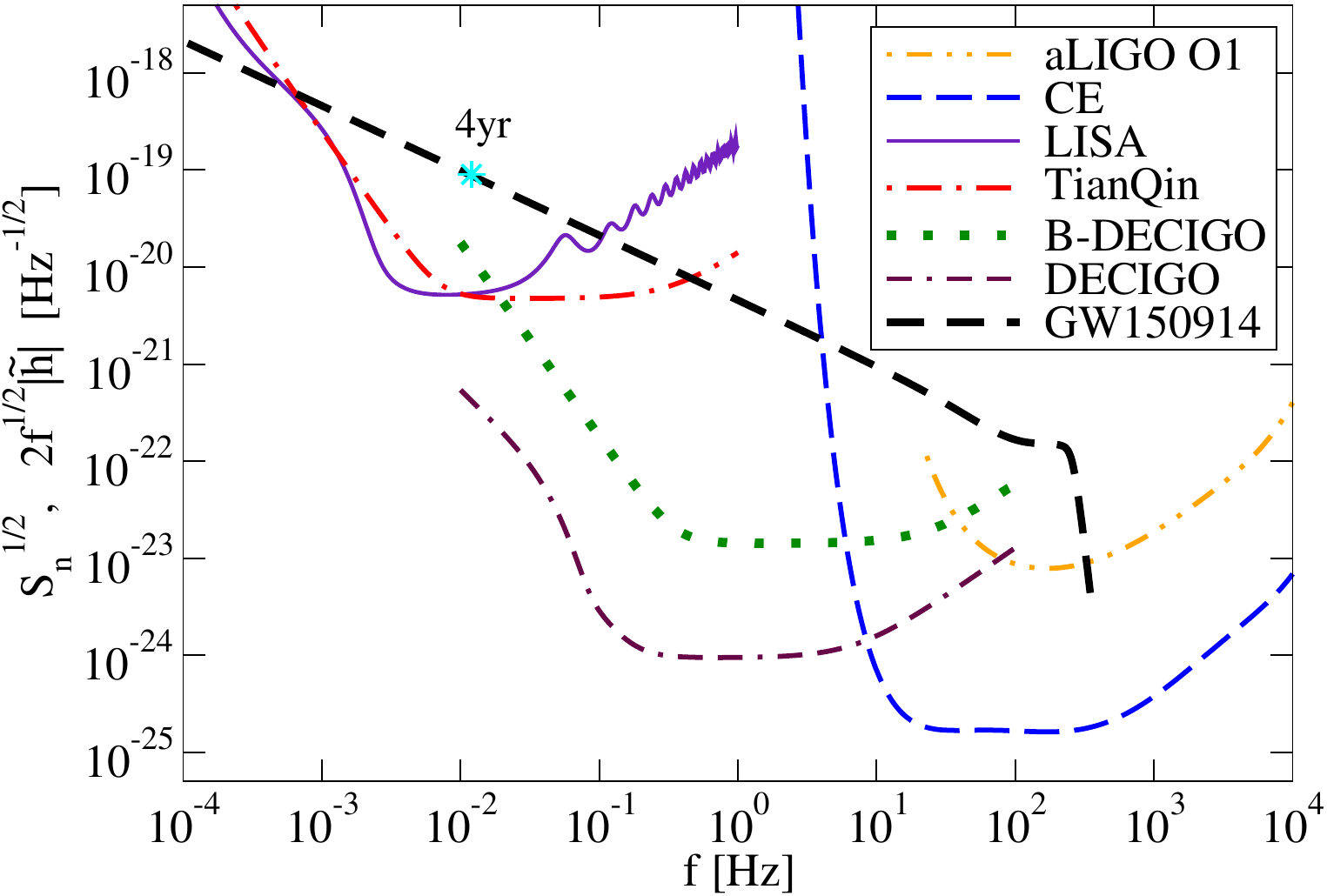}
\caption{
Sensitivities $\sqrt{S_n(f)}$ of various gravitational-wave interferometers.
Also shown is the characteristic amplitude $2\sqrt{f}|\tilde h(f)|$ for GW150914 with 4 years prior to merger displayed as a cyan star.
Observe how the early inspiral portion of the coalescence is observed by space-based detectors, while the late inspiral and merger-ringdown are observed by the ground-based detectors.
}\label{fig:detectors}
\end{center}
\end{figure}

With such an overwhelming success on the GW observational front, many future ground- and space-based detectors have been proposed, planned, and even funded. 
Among these are several upgrades to the current advanced LIGO design~\cite{Ap_Voyager_CE}, along with third generation ground-based detectors Cosmic Explorer (CE)~\cite{Ap_Voyager_CE} and Einstein Telescope (ET)~\cite{ET}, and space-based detectors TianQin~\cite{TianQin}, LISA~\cite{LISA}, B-DECIGO~\cite{B-DECIGO} and DECIGO~\cite{DECIGO,Takahiro} (Fig.~\ref{fig:detectors}).
With roughly 100 times the improvement in sensitivity compared to the current LIGO interferometers, CE will have the ability to stringently constrain modified theories of gravity which are prevalent at high ($1-10^4$ Hz) frequencies (high velocity binaries)~\cite{Yunes:2013dva,Chamberlain:2017fjl}.
On the other side, space-based detectors are sensitive to the low frequency ranges of $10^{-4}-1$ Hz, effectively probing modified theories which are dominant at lower velocities or with larger masses~\cite{Berti:Fisher,Yagi:2009zm,Gair:2012nm,Chamberlain:2017fjl}.

Soon after the discovery of GW150914, Sesana~\cite{Sesana:2016ljz} pointed out that GWs from GW150914-like events are detectable in the future with both LISA and ground-based detectors (Fig.~\ref{fig:detectors}), with expected event rates ranging from 1 to 100 Gpc$^{-3}$yr$^{-1}$~\cite{Gerosa:2019dbe,Sesana:2016ljz}. 
First observed by space-based telescopes in their early inspiral stage, these systems continue to inspiral after leaving the space-band at $1$ Hz for several months before entering CE's band to finally merge at $\sim300$ Hz. 
LISA will be able to give alert to ground-based detectors (allowing for optimizations of ground-based detectors, which can be used to improve upon tests of GR~\cite{Tso:2018pdv}) and electromagnetic telescopes~\cite{Sesana:2016ljz}, while ground-based detectors will help LISA to lower the detection threshold signal-to-ratio (SNR) and enhance the number of detections~\cite{Moore:2019pke,Wong:2018uwb,Cutler:2019krq}.
For GW150914-like events, the SNR can be computed to be $3000$, $9$, $11$, $600$, and $15000$ (as described in the following section) for CE, LISA, TianQin, B-DECIGO, and DECIGO, all above the threshold values (9 for space-based detectors in conjunction with ground-based observations), and therefore observable by the detectors considered in this analysis.
Such \emph{multi-band} GW observations will improve measurement accuracy of binary parameters such as masses and sky positions~\cite{Nair:2015bga,Nair:2018bxj,Vitale:2016rfr,Cutler:2019krq}. 
Multi-band GW astronomy is also possible for more massive binary BHs~\cite{AmaroSeoane:2009ui,Cutler:2019krq} and binary neutron stars~\cite{Isoyama:2018rjb}.

In this letter, we study the impact of multi-band GW astronomy on tests of GR. Such a question was first addressed in~\cite{Barausse:2016eii} for a specific type of non-GR modifications due to radiation of a scalar field using aLIGO+LISA. We here extend this by considering (i) parameterized tests of GR following~\cite{Yunes:2009ke,Yunes_ModifiedPhysics} (see~\cite{Vitale:2016rfr} for a brief work related to this), (ii) various space-borne GW detector combinations with CE and (iii) applications to parity-violating gravity. We also investigate consistency tests of the inspiral and merger-ringdown parts of the waveform~\cite{Ghosh_IMRcon,Ghosh_IMRcon2,Abbott_IMRcon2,Abbott_IMRcon} with multi-band GW observations. Both types of tests have been performed on the observed GW events by the LIGO and Virgo Collaborations (LVC)~\cite{Abbott_IMRcon2,Abbott_IMRcon}.


\label{sec:modified}
\emph{Parameterized tests of GR.}--- Let us begin by considering modifications to GR which violate various fundamental pillars of Einstein's theory.
While one strives to be agnostic towards the list of modified theories of gravity available, a generic formalism of categorizing and constraining them is necessary.
We here consider the parameterized post-Einsteinian (ppE) formalism~\cite{Yunes:2009ke}, which expands the GR gravitational waveform to allow for non-GR variations in the inspiral portion of the waveform phase in the frequency domain\footnote{A slightly different formalism used by the LVC has a one-to-one mapping with the ppE formalism in the inspiral part of the waveform~\cite{Yunes_ModifiedPhysics}.}:
\begin{equation}
\Psi(f) = \Psi_{\text{GR}}(f)(1+\beta u^{2n-5}).
\end{equation}
Here $\Psi_{\text{GR}}$ is the phase in GR, $f$ is the GW frequency, $u=(\pi \mathcal{M} f)^{1/3}$ is the effective relative velocity of binary constituents with chirp mass $\mathcal M = (m_1^3m_2^3/M)^{1/5}$, individual masses $m_i$, and total mass $M\equiv m_1+m_2$. The index $n$ categorizes the \emph{post-Newtonian} (PN) order\footnote{A term of $n$-PN order is proportional to $(u/c)^{2n}$ relative to the leading-order term in the waveform.} at which a given non-GR effect enters the waveform and $\beta$ describes the overall magnitude of such a modification.
Expressions and values of $(\beta,n)$ for specific non-GR theories can be found e.g. in~\cite{Tahura_GdotMap}.

\begin{figure}[htb]
\begin{center}
\includegraphics[width=.6\linewidth]{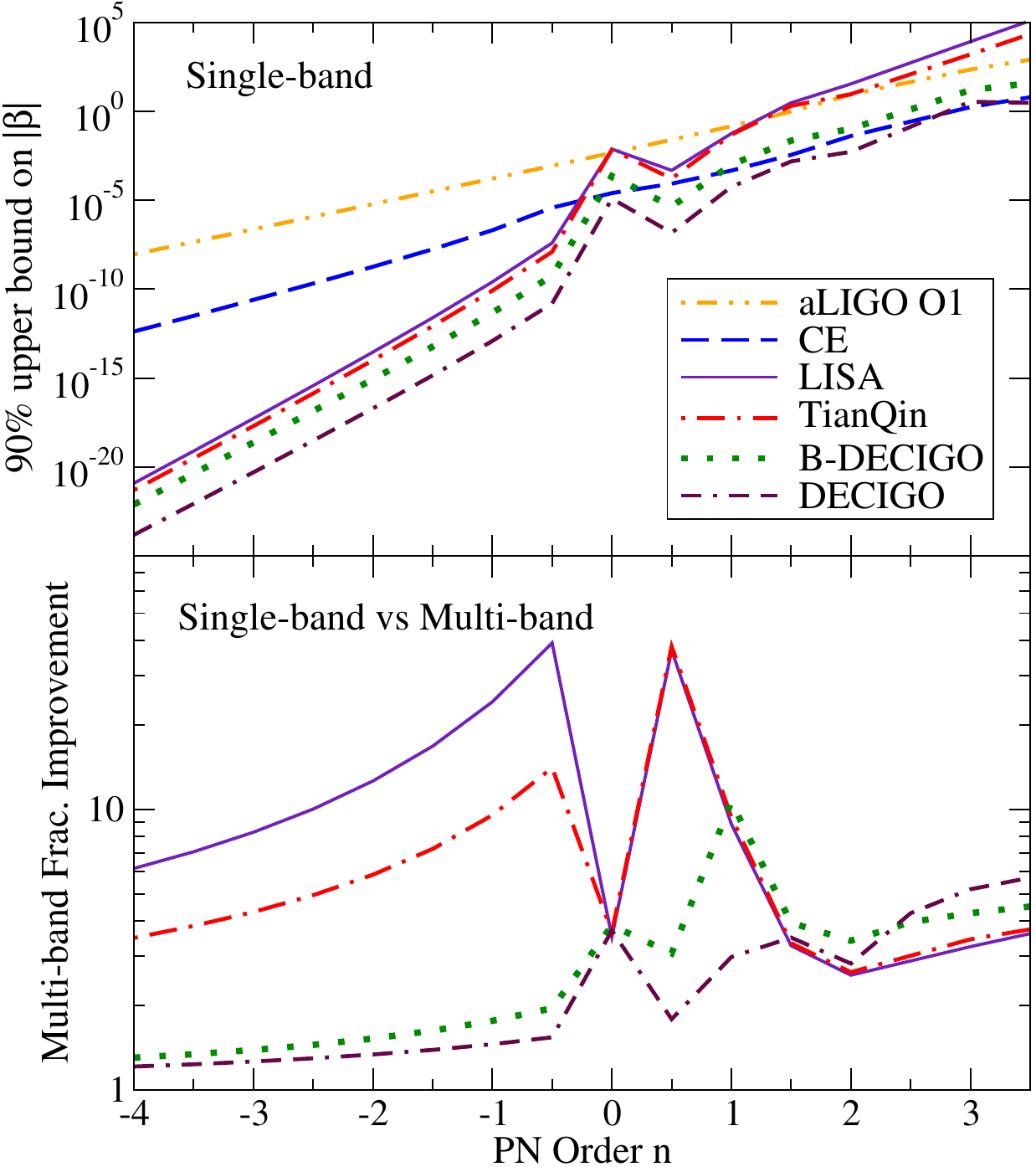}
\caption{
(top) 90\% confidence constraints on the generalized non-GR parameter $|\beta|$ as a function of PN order for GW150914-like events observed on various space- and ground-based detectors individually.
(bottom) Fractional improvement of observations made on the combination of 
CE and space-based detectors (multi-band detection), relative to observations made on CE or space detectors alone (whichever gives stronger bounds).
}\label{fig:betaBounds}
\end{center}
\end{figure}

We utilize a Fisher analysis~\cite{Cutler:Fisher} to obtain constraints on parameters such as $\beta$ from GW150914-like events. 
Such an analysis can be used to reliably approximate results from the more reliable (and computationally-expensive) Bayesian analysis when the SNR is sufficiently high.
For example, the event in question, GW150914, was observed on the O1 detector with an SNR of 25.1, while the CE, LISA, TianQin, B-DECIGO, and DECIGO detectors can be found to observe the same event with SNRs of 3000, 9, 11, 600, and 15000.
Such SNRs are obtained via the expression
\begin{equation}
\rho=\sqrt{(h|h)},
\end{equation}
where $h$ is the gravitational waveform, and the inner product $(a|b)$ is defined as
\begin{equation}
(a|b) \equiv 2 \int\limits^{f_{\text{high}}}_{f_{\text{low}}}\frac{\tilde{a}^*\tilde{b}+\tilde{b}^*\tilde{a}}{S_n(f)}df.
\end{equation}
In the above equation, $S_n(f)$ is the noise spectral density of the given detector, $\tilde{a}$, $\tilde{b}$ are the Fourier-transformed components, and $f_{\text{low,high}}$ are the minimum/maximum frequency cutoffs, dependent on the detector considered.
A complementary analysis by the authors of Ref.~\cite{Yunes_ModifiedPhysics} has proven to find a strong agreement between Fisher and Bayesian analyses on GW150914 (SNR$\sim$24) with the ppE parameter $\beta$ included.
Thus, we expect the Fisher analysis presented here with LISA and TianQin (SNR$\sim$10) to give a valid order-of-magnitude estimate of constraints. We also note that a Fisher analysis was used in e.g.~\cite{Cutler:2019krq} to estimate the measurability of the masses of multi-band sources where the SNR for LISA is only 5.5. We leave a further comparison between Fisher and Bayesian analyses for future work.

We now briefly describe the Fisher analysis process utilized in this investigation.
In such a Fisher analysis, the resulting posterior distribution on the binary parameters $\theta^a$ is given to be Gaussian with root-mean-square errors $\Delta \theta^a = \sqrt{( \tilde\Gamma^{-1} )^{aa}}$, where, assuming Gaussian prior distributions with root-mean-square estimates of parameters $\sigma^{(0)}_{\theta^a}$, the effective Fisher matrix is defined as~\cite{Cutler:Fisher,Poisson:Fisher,Berti:Fisher}
\begin{equation}
\tilde \Gamma_{ij} \equiv \Gamma_{ij} +\frac{1}{\left(\sigma^{(0)}_{\theta^a}\right)^2}\delta_{ij}.
\end{equation}
$\Gamma_{ab}$ is the Fisher information matrix given by
\begin{equation}
\Gamma_{ij}\equiv(\partial_i h | \partial_j h),
\end{equation}
where the gravitational waveform derivatives in question were performed analytically using computer algebra software Mathematica.
Finally, upon the multi-band consideration of such an analysis with resulting Fisher matrices $\Gamma_{ij}^{\text{space}}$ and $\Gamma_{ij}^{\text{ground}}$ from space- and ground-based detectors respectively, the combined Fisher matrix $\tilde \Gamma_{ij}^{\text{total}}$ is computed as 
\begin{equation}
\tilde\Gamma_{ij}^{\text{total}}=\Gamma_{ij}^{\text{space}}+\Gamma_{ij}^{\text{ground}}+\frac{1}{(\sigma^0_{\theta^a})^2}\delta_{ij}.
\end{equation}

In the following analysis, we utilize the sky-averaged ``IMRPhenomD" GR waveform~\cite{PhenomDI,PhenomDII}, which is parameterized in terms of the BH masses $m_i$ and spins $\chi_i$, the time $t_c$ and phase $\phi_c$ at coalescence, and the luminosity distance $D_L$ to the event. 
Therefore, the template parameters included in our Fisher analysis can be written as follows
\begin{equation}
\theta^a=\left( \ln{A}, \phi_c, t_c, \ln{\mathcal{M}_z}, \ln{\eta}, \chi_s, \chi_a, \beta \right),
\end{equation}
where $A\equiv \mathcal{M}^{5/6}/(\sqrt{30} \pi^{2/3} D_L)$ is the generalized amplitude with redshifted chirp mass $\mathcal{M}_z\equiv M\eta^{-3/5}$, total mass $M\equiv m_1+m_2$, symmetric mass ratio $\eta\equiv m_1 m_2/M$, redshift $z$, and symmetric/anti-symmetric combinations of spins $\chi_{s,a}\equiv\frac{1}{2}(\chi_1 \pm \chi_2)$.
Additionally, we assume Gaussian prior distributions on individual BH spins such that $|\chi_i|<1$.
For space-based detectors, we assume that the GW observations on GW150914-like events begin 4 years prior to the BH merger event.

The top panel of Fig.~\ref{fig:betaBounds} displays the corresponding 90\% confidence interval constraints (i.e. the $1.645\sigma$ interval such that one could expect 90\% of the interval estimates to include the given parameter) on $\beta$ as a function of PN order for GW150914-like events\footnote{We choose fiducial values for dimensionless spins of the BHs to be 0. Same choice was made for the inspiral-merger-ringdown tests described later.} observed on each of the ground- and space-based detectors.
We observe that the ground-based detectors are most proficient at probing positive PN orders (corresponding to relatively high-velocity, high-frequency effects), and the space-based detectors are effective at probing negative PN-orders (relatively low-velocity, low-frequency effects). 
The O1 bound is taken from~\cite{Yunes_ModifiedPhysics}, and the CE and LISA bounds are consistent with~\cite{Chamberlain:2017fjl}. The LISA and TianQin bounds are almost identical at positive PN orders because they are dominated by their spin priors.

The bottom panel of Fig.~\ref{fig:betaBounds} displays the fractional improvement made upon a multi-band GW detection with each space-based detector plus CE over a single-band detection, corresponding to:
\begin{equation}
\text{(Fractional Improvement)} \equiv \frac{\min \left( \beta^\mathrm{(CE)},\beta^\mathrm{(space)} \right)}{\beta^\mathrm{(CE+space)}}.
\end{equation}
Observe that multi-band detections can have an improvement by a factor of $\sim 40$ at most, especially for LISA and TianQin.


\label{sec:dCS}
\emph{Application to parity-violating gravity}--- 
We now show the impact of the above improvement in multi-band GW tests of gravity on probing the fundamental pillars of GR. 
To put this into context, we focus on parity invariance in GR and study a string-inspired theory called dynamical Chern-Simons (dCS) gravity~\cite{Jackiw:2003pm,Alexander_cs} which breaks parity in the gravity sector. 
This theory contains one coupling constant $\alpha$ which has the units of length squared and controls the amount of parity violation. 
The correction to the waveform enters at 2PN order and the expression for $\beta$ is given in Eq.~(2) of~\cite{Nair_dCSMap}. Such an expression is derived under the \emph{small coupling approximation}~\cite{Yagi:2012vf}, which assumes that the parity-violation correction is always smaller than the GR contribution and can be treated as a small perturbation. 
This approximation is valid only when the dimensionless coupling constant $\zeta \equiv 16\pi\alpha^2/M^4$ satisfies $\zeta \ll 1$~\cite{Yagi:2012vf}. So far, meaningful bounds have not been placed on this theory from the observed GW events~\cite{Yunes_ModifiedPhysics,Nair_dCSMap}.

\begin{figure}[htb]
\begin{center}
\includegraphics[width=.65\linewidth]{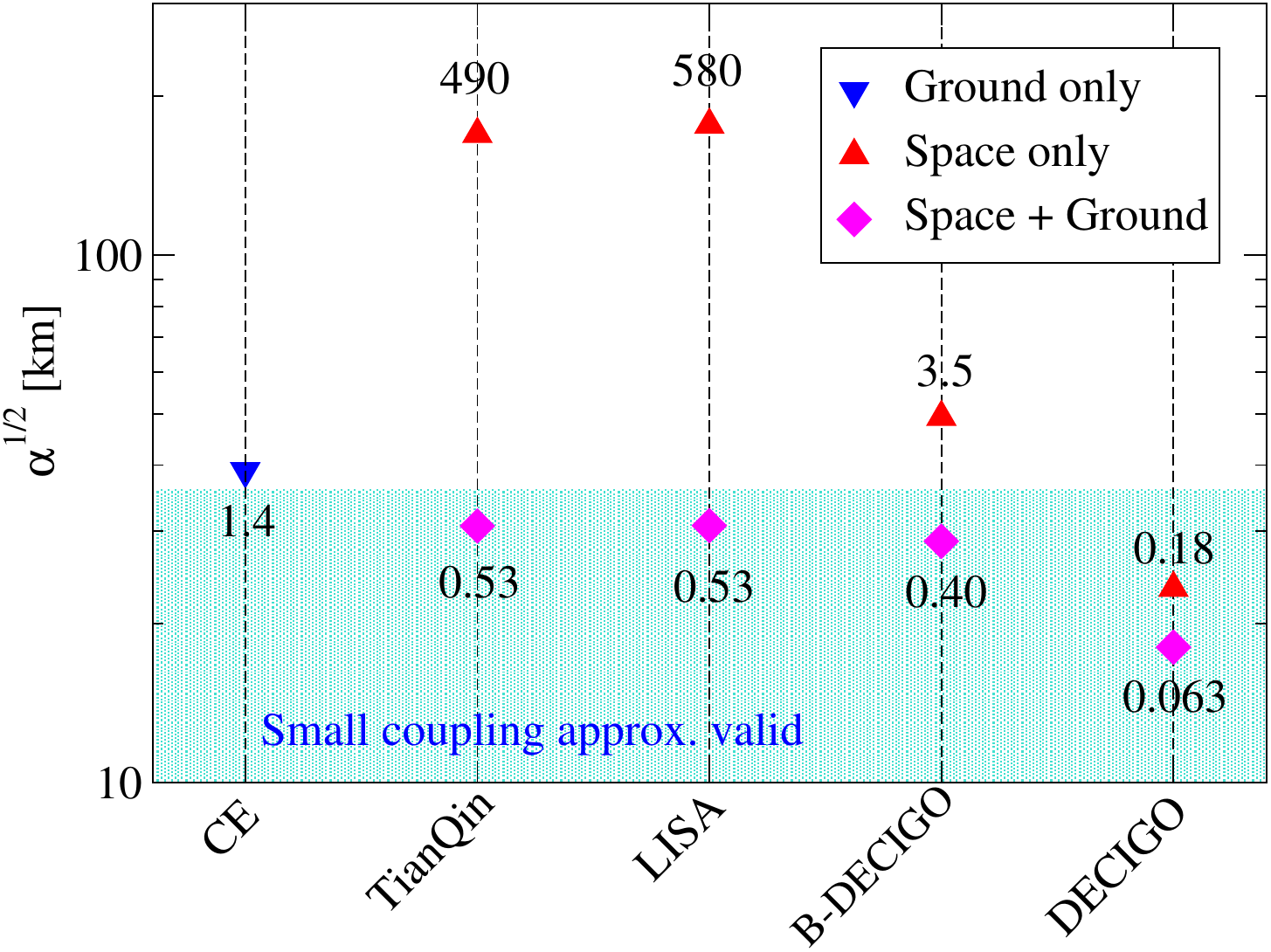}
\caption{
Example 90\% confidence constraints on the parity-violation parameter with CE alone (blue triangle), space-based detectors alone (red triangles) and multi-band space + CE detections (magenta diamonds). 
The number associated to each datum shows the dimensionless parity-violation parameter $\zeta$, and the small coupling approximation is valid only when the bounds fall within the cyan shaded region.
Observe how, for many detectors, such approximation is violated until the multi-band observation is realized. 
For the latter, valid bounds are $\sim 7$ orders of magnitude stronger than the current constraints~\cite{AliHaimoud_dCS,Yagi_dCS}.
}
\label{fig:dCS}
\end{center}
\end{figure}

Figure~\ref{fig:dCS} displays bounds on $\sqrt{\alpha}$ for CE alone, space-based detectors alone, and multi-band GW detections. 
dCS corrections arise during the inspiral phase only when the BHs are spinning, and thus, we recompute the bounds in Fig.~\ref{fig:betaBounds} entering at 2PN order with the fiducial dimensionless spins of $(\chi_1,\chi_2) = (0.15,0)$, consistent with the effective spin measurement of GW150914.
Observe how constraints placed with space- or ground-based detectors alone violate the small-coupling approximation (except for the case of DECIGO) and can place no valid bounds on $\sqrt{\alpha}$~\footnote{The results for CE and DECIGO alone are consistent with those in~\cite{Yagi:2012vf}.}.
However, with multi-band GW observations, the bounds now fall within the validity of the small coupling approximation. Such constraints of $\sqrt{\alpha} \sim \mathcal{O}(10)$ km are stronger than the current bounds from solar system~\cite{AliHaimoud_dCS} and table-top~\cite{Yagi_dCS} experiments by roughly seven orders of magnitude.


\label{sec:IMRDconsistency}
\emph{Inspiral-merger-ringdown consistency tests.}---  So far, we have focused on probing non-GR corrections entering in the inspiral, but one can also test the consistency between the inspiral and merger-ringdown parts of the waveform assuming GR is correct~\cite{Ghosh_IMRcon,Ghosh_IMRcon2,Ghosh_2017,Abbott_IMRcon,Abbott_IMRcon2}, as follows.
Through use of the numerical relativity fits found in Ref.~\cite{PhenomDII}, the remnant BH mass $M_f$ and spin $\chi_f$ can be predicted entirely from the individual masses $m_i$ and spins $\chi_i$ prior to the merger. 
Thus, one can first estimate these parameters independently from both the inspiral and merger-ringdown of the waveform using the GR template, and next check the consistency between the two.
If statistically significant deviations between the two were observed, evidence could be presented for deviations from GR~\cite{Ghosh_IMRcon}.
We here demonstrate how one can improve the discriminatory power to detect deviations from GR with multi-band GW astronomy.

We compute the probability distribution of $M_f$ and $\chi_f$ as follows.
We begin by using Fisher-analysis methods~\cite{Cutler:Fisher} to estimate the four-dimensional Gaussian posterior probability distributions $P_{\text{I,MR}}(m_1,m_2,\chi_1,\chi_2)$ from the observed inspiral (I) and merger-ringdown (MR) signals independently, with the transition frequency between the two defined to be $f_{\text{trans}}=132$ Hz for GW150914-like events~\cite{Abbott_IMRcon}.
This is done after marginalizing over all other binary parameters present in the template waveform\footnote{Marginalization over a given parameter is typically accomplished by integration over the full range of values, or in the case of multi-variate Gaussian distributions by simply removing the corresponding row and column from the covariance matrix $\Sigma_{ij}\equiv\Gamma_{ij}^{-1}$.}.
Through the Jacobian transformation matrix and the numerical relativity fits for the remnant mass $M_f(m_1,m_2,\chi_1,\chi_2)$ and spin $\chi_f(m_1,m_2,\chi_1,\chi_2)$~\cite{PhenomDII}, such posterior distributions may be transformed into $P_{\text{IMR}}(M_f,\chi_f)$, $P_{\text{I}}(M_f,\chi_f)$, and $P_{\text{MR}}(M_f,\chi_f)$, all of which must overlap in the $(M_f,\chi_f)$ plane if the GR assumption is correct.

Typically, agreement between the above distributions is measured by transforming the posteriors once again into the single probability distribution $P \left( \Delta M_f/\bar{M}_f, \Delta \chi_f/\bar{\chi}_f \right)$ following Eq. (A.2) of Ref.~\cite{Ghosh_2017}. 
Here, $\Delta M_f \equiv M^{\text{I}}_f-M^{\text{MR}}_f$ and $\Delta \chi_f \equiv \chi^{\text{I}}_f-\chi^{\text{MR}}_f$ describe the differences in the GR predictions of final mass and spin between inspiral and merger-ringdown, while $\bar{M}_f\equiv(M^{\text{I}}_f+M^{\text{MR}}_f)/2$ and $\bar{\chi}_f\equiv(\chi^{\text{I}}_f+\chi^{\text{MR}}_f)/2$ are computed from the averages between the two.
Finally,  agreement of such a posterior with the GR value of $\left( \Delta M_f/\bar{M}_f, \Delta \chi_f/\bar{\chi}_f \right) \big|_{\text{GR}} \equiv (0,0)$ can determine the consistency of the GW signal with GR.

While Fisher analyses can not predict central values like the more comprehensive Bayesian analysis used in Refs.~\cite{Abbott_IMRcon,Abbott_IMRcon2,Ghosh_IMRcon,Ghosh_IMRcon2,Ghosh_2017}, they can estimate the \emph{size} of posterior probability distributions, which is of high value in the interest of estimating future discriminatory power from the GR value of $\left( \Delta M_f/\bar{M}_f, \Delta \chi_f/\bar{\chi}_f \right) = (0,0)$.
In particular, we consider the \emph{area} of the 90\% confidence region as a metric towards comparing the resolving power of this test upon use of future detectors, and combinations thereof.

\begin{figure}[htb]
\begin{center}
\includegraphics[width=.45\linewidth]{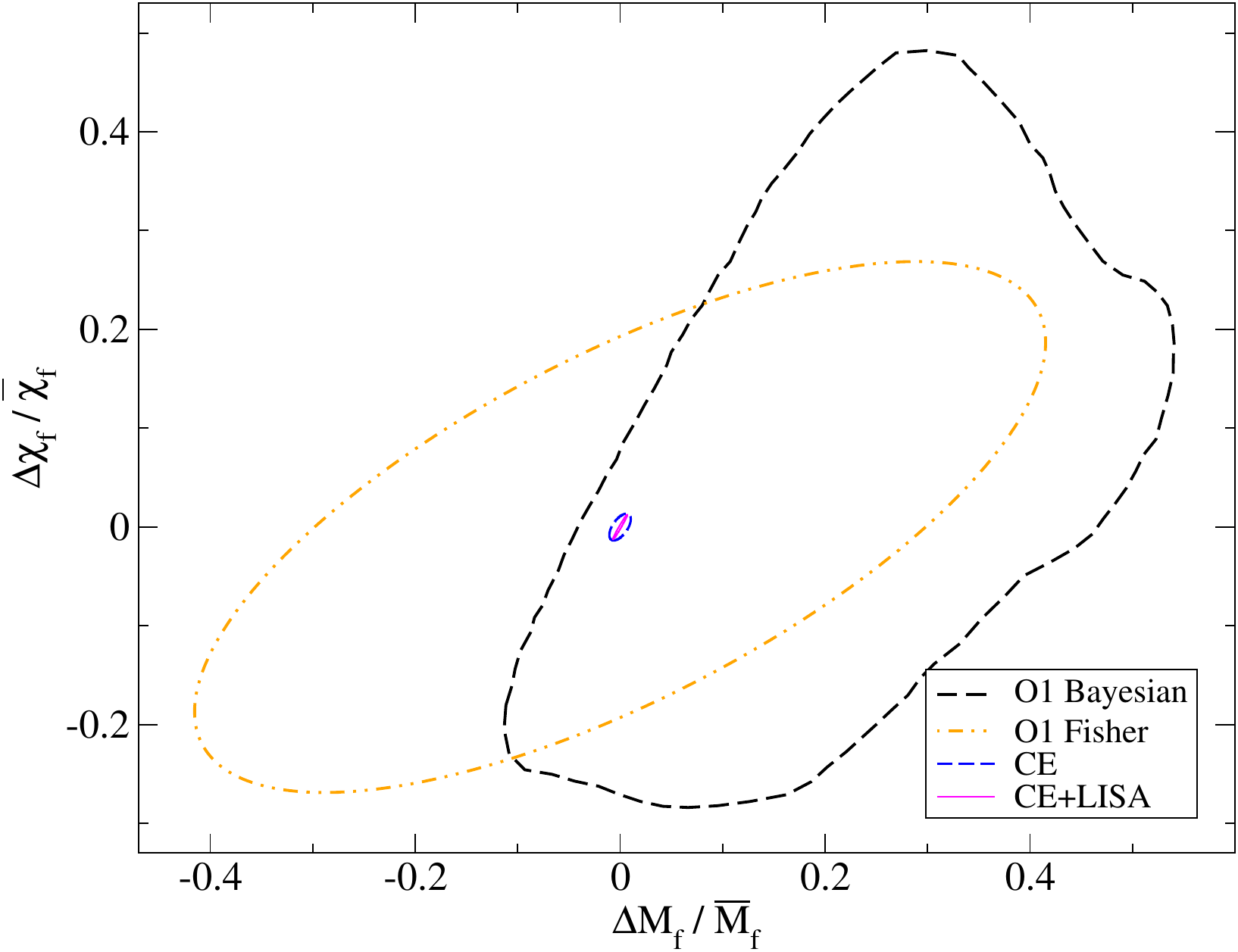}
\includegraphics[width=.49\linewidth]{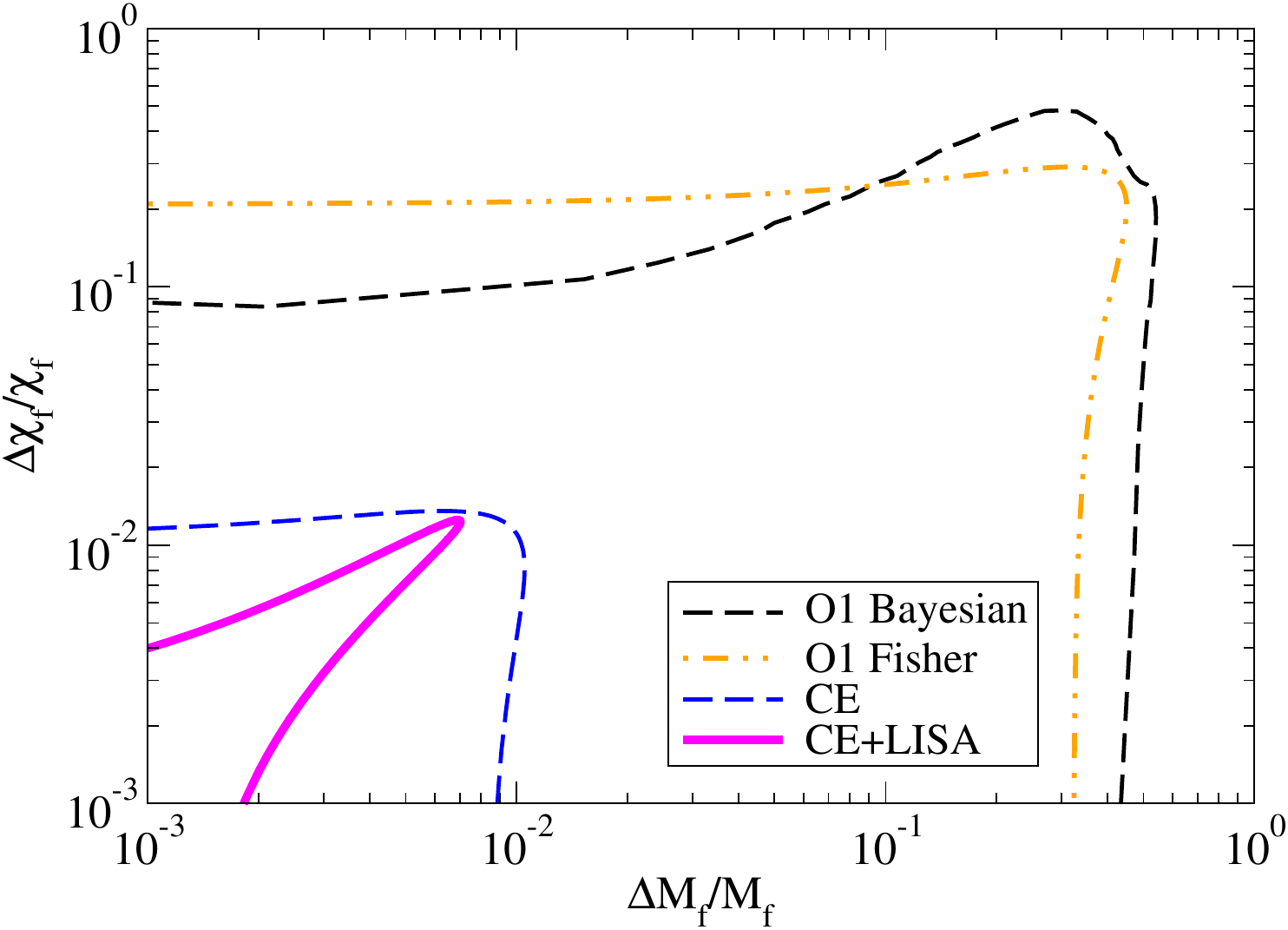}
\caption{
Resulting 90\% confidence ellipses (presented in both linear-space (left) and log-space (right) for demonstration) from the $(\Delta M_f/\bar{M}_f,\Delta\chi_f/\bar{\chi}_f)$ posterior probability distribution using single-band CE observations and multi-band observations with both CE and LISA.
The value consistent with GR corresponds to $(\Delta M_f/\bar{M}_f,\Delta\chi_f/\bar{\chi}_f )= (0,0)$.
Also shown for comparison is the aLIGO O1 result found with the full Bayesian analysis of Ref.~\cite{Abbott_IMRcon}, agreeing with the Fisher-estimates here within 10\% of the total area of the 90\% contours.
The area of such confidence regions (summarized in Table~\ref{tab:IMRDconsistency}) is indicative of the effective modified gravity resolving power, and can be seen to improve by $\sim7-10$ times upon the multi-band observation as opposed to CE alone.
}\label{fig:IMRDconsistency}
\end{center}
\end{figure}

\begin{table}[htb]
\centering
\begin{tabular}{c|c}
Detector & 90\% area \\
\hline
\hline
LIGO O1 (Fisher) & 0.25 \\
LIGO O1 (Bayesian)~\cite{Abbott_IMRcon} & 0.29 \\
CE & $3.6\times 10^{-4}$\\
\hline
TianQin+CE & $4.9\times 10^{-5}$\\
LISA+CE & $5.1\times 10^{-5}$\\
B-DECIGO+CE & $4.3\times 10^{-5}$\\
DECIGO+CE & $3.8\times 10^{-5}$\\
\end{tabular}
\caption{
Resulting areas of the 90\% confidence ellipses from the $(\Delta M_f/\bar{M}_f,\Delta\chi_f/\bar{\chi}_f)$ posterior distributions for GW150914-like events found in Fig.~\ref{fig:IMRDconsistency}.
}\label{tab:IMRDconsistency}
\end{table}

Figure~\ref{fig:IMRDconsistency} displays the results of the IMR consistency test for GW150914-like events, showing the $(\Delta M_f/\bar{M}_f,\Delta\chi_f/\bar{\chi}_f)$ posterior 90\% confidence regions for LIGO O1 (extracted results from the \texttt{IMRPhenomPv2} Bayesian results of Ref.\footnote{Similar results were found with the non-precessing \texttt{SEOBNRv4} model in Ref.~\cite{Abbott_IMRcon}.}~\cite{Abbott_IMRcon} and Fisher for comparison), CE, and the combination of CE and LISA\footnote{The IMR consistency test for stellar-mass BH binaries can not be performed entirely with space-based detectors due to their inability to observe the merger-ringdown portion of the signal (Ref.~\cite{Hughes:2004vw} showed that supermassive BH binaries are compatible with such observations), accomplished proficiently with ground-based detectors.}.
Table~\ref{tab:IMRDconsistency} further summarizes the results by listing the areas of such 90\% confidence regions.
First, observe that the 90\% confidence region areas between the Bayesian~\cite{Abbott_IMRcon} and Fisher analyses for LIGO O1 agree within 10\%, demonstrating good agreement between the two methods.
Second, notice that CE will observe significant reductions in the 90\% contour area by $\sim 3$ orders of magnitude from the current test with O1. Third, observe that multi-band GW observations will further improve the consistency test by a factor of $7-10$ compared to single-band measurements with CE alone.
Such an improvement in the size of posterior probability distributions for multi-band GW observations can effectively allow one to discriminate non-GR effects that might not be visible when observing with ground- or space-based detectors alone.
The fact that all multi-band choices show similar results suggest that the error is mostly dominated by the merger-ringdown measurement from CE.


\label{sec:discussion}
\emph{Conclusion.}--- In this letter, we have demonstrated the power in making multi-band observations of GWs, specifically for tests of gravity.
We first considered parameterized tests of GR and found that multi-band GW observations improve the bounds on non-GR generic parameters up to $\sim 40$ times compared to either ground- or space-based detectors alone. 
We then applied this result to parity-violating gravity and found that it is crucial to realize such multi-band observations to place meaningful bounds in this theory. 
Such constraints are stronger than the current bounds by seven orders of magnitude. 
Finally, we studied the consistency between the inspiral and merger-ringdown parts of the waveform in GR.
We found up to an order-of-magnitude improvement in such tests upon the use of the combination of space- and ground-based detectors, rather than using ground ones alone.
This highlights the advantages of multi-band GW astronomy with the highly enhanced opportunity to shed light on even the most minute deviations from GR in the extreme gravity sector.

Future work in this direction can enrich the current analysis by simulating the multi-band event rates described in Refs.~\cite{Gerosa:2019dbe,Sesana:2016ljz}, allowing one to ``stack" multiple events and further improve our estimated constraints.
In addition, modified theories of gravity that alter the GW amplitude rather than the phase may be considered~\cite{Alexander:2007kv,Yunes:2008bu,Yunes:2010yf,Yagi:2017zhb}.
We used a Fisher analysis though such an analysis is known to have fallbacks, especially when the SNR is low~\cite{Gair:2012nm,Stevenson:2015bqa,Vallisneri,Rodriguez:2013mla}.
One can also repeat the presented analysis with a Bayesian approach rather than the Fisher one considered here, in order to confirm the results presented here.
We conclude by referring the readers to the upcoming work~\cite{Carson_multiBandPRD} with a more detailed discussion of our analysis presented here with a comprehensive list of future bounds on modified theories of gravity with multi-band GW astronomy.


\emph{Additional note.}--- A complementary analysis with similar conclusions was submitted to arXiv shortly after the release of this letter~\cite{Gnocchi:2019jzp}. While finalizing this work, Ref.~\cite{Moore:2019pke} was submitted to arXiv which reduced the expected number of events, though one can use the information from ground-based detectors to still detect $\mathcal{O}(1)$ events.

\label{acknowledgments}
\emph{Acknowledgments.}--- We thank Katerina Chatziioannou and Carl-Johan Haster for providing valuable comments on the comparisons between Fisher and Bayesian analyses. 
We also thank Enrico Barausse and Jian-dong Zhang for pointing us to the correct noise curve for TianQin.
Additionally, we thank Nathan Johnson-McDaniel for many insightful comments on the IMR consistency tests.
Finally, we acknowledge Takahiro Tanaka and Davide Gerosa for pointing out recent related works.
Z.C. and K.Y. acknowledge support from NSF Award PHY-1806776. 
K.Y. would like to also acknowledge support by the COST Action GWverse CA16104 and JSPS KAKENHI Grants No. JP17H06358.

\section*{References}

\bibliographystyle{iopart-num}
\bibliography{Zack}

\newcommand{\noop}[1]{}
\providecommand{\newblock}{}
\begin{thebibliography}{10}
\expandafter\ifx\csname url\endcsname\relax
  \def\url#1{{\tt #1}}\fi
\expandafter\ifx\csname urlprefix\endcsname\relax\def\urlprefix{URL }\fi
\providecommand{\eprint}[2][]{\url{#2}}

\bibitem{Will_SolarSystemTest}
Will C~M 2014 {\em Living Reviews in Relativity\/} {\bf 17} 4 ISSN 1433-8351
  \urlprefix\url{https://doi.org/10.12942/lrr-2014-4}

\bibitem{Stairs_BinaryPulsarTest}
Stairs I~H 2003 {\em Living Rev. Rel.\/} {\bf 6} 5 (\textit{Preprint}
  \eprint{astro-ph/0307536})

\bibitem{Wex_BinaryPulsarTest}
Wex N 2014  (\textit{Preprint} \eprint{1402.5594})

\bibitem{Ferreira_CosmologyTest}
Ferreira P~G 2019  (\textit{Preprint} \eprint{1902.10503})

\bibitem{Clifton_CosmologyTest}
Clifton T, Ferreira P~G, Padilla A and Skordis C 2012 {\em Phys. Rept.\/} {\bf
  513} 1--189 (\textit{Preprint} \eprint{1106.2476})

\bibitem{Joyce_CosmologyTest}
Joyce A, Jain B, Khoury J and Trodden M 2015 {\em Phys. Rept.\/} {\bf 568}
  1--98 (\textit{Preprint} \eprint{1407.0059})

\bibitem{Koyama_CosmologyTest}
Koyama K 2016 {\em Rept. Prog. Phys.\/} {\bf 79} 046902 (\textit{Preprint}
  \eprint{1504.04623})

\bibitem{Salvatelli_CosmologyTest}
Salvatelli V, Piazza F and Marinoni C 2016 {\em JCAP\/} {\bf 1609} 027
  (\textit{Preprint} \eprint{1602.08283})

\bibitem{GW150914}
Abbott B~P {\em et~al.\/} (LIGO Scientific, Virgo) 2016 {\em Phys. Rev.
  Lett.\/} {\bf 116} 241102 (\textit{Preprint} \eprint{1602.03840})

\bibitem{Abbott_IMRcon2}
Abbott B~P {\em et~al.\/} (LIGO Scientific, Virgo) 2016 {\em Phys. Rev.
  Lett.\/} {\bf 116} 221101 [Erratum: Phys. Rev. Lett.121,no.12,129902(2018)]
  (\textit{Preprint} \eprint{1602.03841})

\bibitem{Yunes_ModifiedPhysics}
Yunes N, Yagi K and Pretorius F 2016 {\em Phys. Rev. D\/} {\bf 94}(8) 084002
  \urlprefix\url{https://link.aps.org/doi/10.1103/PhysRevD.94.084002}

\bibitem{GW_Catalogue}
Abbott B~P {\em et~al.\/} (LIGO Scientific, Virgo) 2018  (\textit{Preprint}
  \eprint{1811.12907})

\bibitem{TheLIGOScientific:2017qsa}
Abbott B~P {\em et~al.\/} (Virgo, LIGO Scientific) 2017 {\em Phys. Rev.
  Lett.\/} {\bf 119} 161101 (\textit{Preprint} \eprint{1710.05832})

\bibitem{Abbott_IMRcon}
Abbott B~P {\em et~al.\/} (LIGO Scientific, Virgo) 2019  (\textit{Preprint}
  \eprint{1903.04467})

\bibitem{Monitor:2017mdv}
Abbott B~P {\em et~al.\/} (LIGO Scientific, Virgo, Fermi-GBM, INTEGRAL) 2017
  {\em Astrophys. J.\/} {\bf 848} L13 (\textit{Preprint} \eprint{1710.05834})

\bibitem{Abbott:2018lct}
Abbott B~P {\em et~al.\/} (LIGO Scientific, Virgo) 2018  (\textit{Preprint}
  \eprint{1811.00364})

\bibitem{Ap_Voyager_CE}
Ligo-t1400316-v4: Instrument science white paper
  \url{https://dcc.ligo.org/ligo-T1400316/public}

\bibitem{ET}
The {ET} project website \url{http://www.et-gw.eu/}

\bibitem{TianQin}
Shi C, Bao J, Wang H, Zhang J~d, Hu Y, Sesana A, Barausse E, Mei J and Luo J
  2019  (\textit{Preprint} \eprint{1902.08922})

\bibitem{LISA}
Robson T, Cornish N and Liu C 2019 {\em Class. Quant. Grav.\/} {\bf 36} 105011
  (\textit{Preprint} \eprint{1803.01944})

\bibitem{B-DECIGO}
Isoyama S, Nakano H and Nakamura T 2018 {\em PTEP\/} {\bf 2018} 073E01
  (\textit{Preprint} \eprint{1802.06977})

\bibitem{DECIGO}
Yagi K and Seto N 2011 {\em Phys. Rev.\/} {\bf D83} 044011 [Erratum: Phys.
  Rev.D95,no.10,109901(2017)] (\textit{Preprint} \eprint{1101.3940})

\bibitem{Takahiro}
Yagi K and Tanaka T 2010 {\em Prog. Theor. Phys.\/} {\bf 123} 1069--1078
  (\textit{Preprint} \eprint{0908.3283})

\bibitem{Yunes:2013dva}
Yunes N and Siemens X 2013 {\em Living Rev. Rel.\/} {\bf 16} 9
  (\textit{Preprint} \eprint{1304.3473})

\bibitem{Chamberlain:2017fjl}
Chamberlain K and Yunes N 2017 {\em Phys. Rev.\/} {\bf D96} 084039
  (\textit{Preprint} \eprint{1704.08268})

\bibitem{Berti:Fisher}
Berti E, Buonanno A and Will C~M 2005 {\em Phys. Rev.\/} {\bf D71} 084025
  (\textit{Preprint} \eprint{gr-qc/0411129})

\bibitem{Yagi:2009zm}
Yagi K and Tanaka T 2010 {\em Phys. Rev.\/} {\bf D81} 064008 [Erratum: Phys.
  Rev.D81,109902(2010)] (\textit{Preprint} \eprint{0906.4269})

\bibitem{Gair:2012nm}
Gair J~R, Vallisneri M, Larson S~L and Baker J~G 2013 {\em Living Rev. Rel.\/}
  {\bf 16} 7 (\textit{Preprint} \eprint{1212.5575})

\bibitem{Sesana:2016ljz}
Sesana A 2016 {\em Phys. Rev. Lett.\/} {\bf 116} 231102 (\textit{Preprint}
  \eprint{1602.06951})

\bibitem{Gerosa:2019dbe}
Gerosa D, Ma S, Wong K~W~K, Berti E, O'Shaughnessy R, Chen Y and Belczynski K
  2019 {\em Phys. Rev.\/} {\bf D99} 103004 (\textit{Preprint}
  \eprint{1902.00021})

\bibitem{Tso:2018pdv}
Tso R, Gerosa D and Chen Y 2018  (\textit{Preprint} \eprint{1807.00075})

\bibitem{Moore:2019pke}
Moore C~J, Gerosa D and Klein A 2019  (\textit{Preprint} \eprint{1905.11998})

\bibitem{Wong:2018uwb}
Wong K~W~K, Kovetz E~D, Cutler C and Berti E 2018 {\em Phys. Rev. Lett.\/} {\bf
  121} 251102 (\textit{Preprint} \eprint{1808.08247})

\bibitem{Cutler:2019krq}
Cutler C {\em et~al.\/} 2019  (\textit{Preprint} \eprint{1903.04069})

\bibitem{Nair:2015bga}
Nair R, Jhingan S and Tanaka T 2016 {\em PTEP\/} {\bf 2016} 053E01
  (\textit{Preprint} \eprint{1504.04108})

\bibitem{Nair:2018bxj}
Nair R and Tanaka T 2018 {\em JCAP\/} {\bf 1808} 033 [Erratum:
  JCAP1811,no.11,E01(2018)] (\textit{Preprint} \eprint{1805.08070})

\bibitem{Vitale:2016rfr}
Vitale S 2016 {\em Phys. Rev. Lett.\/} {\bf 117} 051102 (\textit{Preprint}
  \eprint{1605.01037})

\bibitem{AmaroSeoane:2009ui}
Amaro-Seoane P and Santamaria L 2010 {\em Astrophys. J.\/} {\bf 722} 1197--1206
  (\textit{Preprint} \eprint{0910.0254})

\bibitem{Isoyama:2018rjb}
Isoyama S, Nakano H and Nakamura T 2018 {\em PTEP\/} {\bf 2018} 073E01
  (\textit{Preprint} \eprint{1802.06977})

\bibitem{Barausse:2016eii}
Barausse E, Yunes N and Chamberlain K 2016 {\em Phys. Rev. Lett.\/} {\bf 116}
  241104 (\textit{Preprint} \eprint{1603.04075})

\bibitem{Yunes:2009ke}
Yunes N and Pretorius F 2009 {\em Phys. Rev.\/} {\bf D80} 122003
  (\textit{Preprint} \eprint{0909.3328})

\bibitem{Ghosh_IMRcon}
Ghosh A {\em et~al.\/} 2016 {\em Phys. Rev.\/} {\bf D94} 021101
  (\textit{Preprint} \eprint{1602.02453})

\bibitem{Ghosh_IMRcon2}
Ghosh A, Johnson-Mcdaniel N~K, Ghosh A, Mishra C~K, Ajith P, Del~Pozzo W, Berry
  C~P~L, Nielsen A~B and London L 2018 {\em Class. Quant. Grav.\/} {\bf 35}
  014002 (\textit{Preprint} \eprint{1704.06784})

\bibitem{Tahura_GdotMap}
Tahura S and Yagi K 2018 {\em Phys. Rev.\/} {\bf D98} 084042 (\textit{Preprint}
  \eprint{1809.00259})

\bibitem{Cutler:Fisher}
Cutler C and Flanagan E~E 1994 {\em Phys. Rev. D\/} {\bf 49}(6) 2658--2697
  \urlprefix\url{https://link.aps.org/doi/10.1103/PhysRevD.49.2658}

\bibitem{Poisson:Fisher}
Poisson E and Will C~M 1995 {\em Phys. Rev. D\/} {\bf 52}(2) 848--855
  \urlprefix\url{https://link.aps.org/doi/10.1103/PhysRevD.52.848}

\bibitem{PhenomDI}
Khan S, Husa S, Hannam M, Ohme F, P\"urrer M, Forteza X~J and Boh\'e A 2016
  {\em Phys. Rev. D\/} {\bf 93}(4) 044007
  \urlprefix\url{https://link.aps.org/doi/10.1103/PhysRevD.93.044007}

\bibitem{PhenomDII}
Husa S, Khan S, Hannam M, P\"urrer M, Ohme F, Forteza X~J and Boh\'e A 2016
  {\em Phys. Rev. D\/} {\bf 93}(4) 044006
  \urlprefix\url{https://link.aps.org/doi/10.1103/PhysRevD.93.044006}

\bibitem{Jackiw:2003pm}
Jackiw R and Pi S~Y 2003 {\em Phys. Rev.\/} {\bf D68} 104012 (\textit{Preprint}
  \eprint{gr-qc/0308071})

\bibitem{Alexander_cs}
Alexander S and Yunes N 2009 {\em Phys. Rept.\/} {\bf 480} 1--55
  (\textit{Preprint} \eprint{0907.2562})

\bibitem{Nair_dCSMap}
Nair R, Perkins S, Silva H~O and Yunes N 2019  (\textit{Preprint}
  \eprint{1905.00870})

\bibitem{Yagi:2012vf}
Yagi K, Yunes N and Tanaka T 2012 {\em Phys. Rev. Lett.\/} {\bf 109} 251105
  [Erratum: Phys. Rev. Lett.116,no.16,169902(2016)] (\textit{Preprint}
  \eprint{1208.5102})

\bibitem{AliHaimoud_dCS}
Ali-Haimoud Y and Chen Y 2011 {\em Phys. Rev.\/} {\bf D84} 124033
  (\textit{Preprint} \eprint{1110.5329})

\bibitem{Yagi_dCS}
Yagi K, Yunes N and Tanaka T 2012 {\em Phys. Rev.\/} {\bf D86} 044037 [Erratum:
  Phys. Rev.D89,049902(2014)] (\textit{Preprint} \eprint{1206.6130})

\bibitem{Ghosh_2017}
Ghosh A, Johnson-McDaniel N~K, Ghosh A, Mishra C~K, Ajith P, Pozzo W~D, Berry
  C~P~L, Nielsen A~B and London L 2017 {\em Classical and Quantum Gravity\/}
  {\bf 35} 014002 \urlprefix\url{https://doi.org/10.1088%2F1361-6382%2Faa972e}

\bibitem{Hughes:2004vw}
Hughes S~A and Menou K 2005 {\em Astrophys. J.\/} {\bf 623} 689--699
  (\textit{Preprint} \eprint{astro-ph/0410148})

\bibitem{Alexander:2007kv}
Alexander S, Finn L~S and Yunes N 2008 {\em Phys. Rev.\/} {\bf D78} 066005
  (\textit{Preprint} \eprint{0712.2542})

\bibitem{Yunes:2008bu}
Yunes N and Finn L~S 2009 {\em J. Phys. Conf. Ser.\/} {\bf 154} 012041
  (\textit{Preprint} \eprint{0811.0181})

\bibitem{Yunes:2010yf}
Yunes N, O'Shaughnessy R, Owen B~J and Alexander S 2010 {\em Phys. Rev.\/} {\bf
  D82} 064017 (\textit{Preprint} \eprint{1005.3310})

\bibitem{Yagi:2017zhb}
Yagi K and Yang H 2018 {\em Phys. Rev.\/} {\bf D97} 104018 (\textit{Preprint}
  \eprint{1712.00682})

\bibitem{Stevenson:2015bqa}
Stevenson S, Ohme F and Fairhurst S 2015 {\em Astrophys. J.\/} {\bf 810} 58
  (\textit{Preprint} \eprint{1504.07802})

\bibitem{Vallisneri}
Vallisneri M 2008 {\em Phys. Rev. D\/} {\bf 77}(4) 042001
  \urlprefix\url{https://link.aps.org/doi/10.1103/PhysRevD.77.042001}

\bibitem{Rodriguez:2013mla}
Rodriguez C~L, Farr B, Farr W~M and Mandel I 2013 {\em Phys. Rev.\/} {\bf D88}
  084013 (\textit{Preprint} \eprint{1308.1397})

\bibitem{Carson_multiBandPRD}
Carson Z and Yagi K \noop{3001}in preparation

\bibitem{Gnocchi:2019jzp}
Gnocchi G, Maselli A, Abdelsalhin T, Giacobbo N and Mapelli M 2019
  (\textit{Preprint} \eprint{1905.13460})

\end{thebibliography}
\end{document}